\documentclass[amsmath, amssymb, onecolumn,aps]{revtex4}
\usepackage{amsthm}
\newcommand{\ra}{\rangle}
\newcommand{\la}{\langle}

\DeclareMathOperator{\Tr}{Tr}
\DeclareMathOperator{\ido}{\openone}
\newtheorem{theorem}{Theorem}[section]
\newtheorem{lem}{Lemma}[section]

\usepackage{graphicx}

\begin{document}
\title{The distillability problem revisited}
\author{Lieven Clarisse}
\email{lc181@york.ac.uk}
\affiliation{Dept. of Mathematics, The University of York, Heslington, York YO10 5DD, U.K.}
\begin{abstract}
An important open problem in quantum information theory is the question of the existence of NPT bound entanglement. In the past years, little progress has been made, mainly because of the lack of mathematical tools to address the problem. 
(i) In an attempt to overcome this, we show how the distillability problem can be reformulated as a special instance of the separability problem, for which a large number of tools and techniques are available. 
(ii) Building up to this we also show how the problem can be formulated as a Schmidt number problem.
(iii) A numerical method for detecting distillability is presented  and strong evidence is given that all 1-copy undistillable Werner states are also 4-copy undistillable.  
(iv) The same method is used to estimate the volume of distillable states, and the results suggest that bound entanglement is primarily a phenomenon found in low dimensional quantum systems.
(v) Finally, a set of one parameter states is presented which we conjecture to exhibit all forms of distillability.
\end{abstract}
\maketitle

\section{Introduction}
Only recently the mathematical definition of \emph{entanglement} in quantum information was \emph{rigorously} supplemented by a physical interpretation.  The definition, as introduced by Werner \cite{Werner89}, is well known.
A bipartite system $\rho \in {\cal L}({\cal H}_A \otimes {\cal H}_B)$ is called separable if and only if $\rho$ can be expanded as
$$
\rho=\sum_i p_i \rho^A_i \otimes \rho^B_i,
$$
with $p_i>0$. If this is not possible, $\rho$ is called entangled. 
When a state cannot be written in a separable form, does that mean it cannot be constructed locally? For a single copy of a state this is obvious \cite{HHH98}, while in the asymptotic regime this was only recently proved \cite{YHHS05}. 

Considerable effort has been devoted in quantifying the amount of entanglement present in a state, usually with the aid of \emph{entanglement measures} \cite{Horodecki01, PV05}. 
The principal physical demands on such an entanglement measure is that it vanishes on separable states, and that it is non-increasing under local operations and classical communication (LOCC). 
However, it is by no means evident what `entanglement present in a state' means.
There are two obvious physical options to this: we can either mean the amount of entanglement used to construct the state or the amount of entanglement we can recover from the state. 
Usually this is done in the asymptotic regime and entanglement is measured with reference to the singlet state $|\Psi\ra=\frac{1}{\sqrt{2}} (|00\ra + |11\ra)$. The two associated entanglement measures are then called the entanglement cost $E_C$ and the distillable entanglement $E_D$. A celebrated result of quantum information is that both values coincide for pure states and are equal to the von Neumann entropy of the reduced density operator \cite{BBPS96}. For mixed states it was shown that any entanglement measure $E$ should satisfy $E_D \leq E \leq E_C$ \cite{HHH99}. The results of Ref.\ \cite{YHHS05} we mentioned earlier implies that $E_C>0$ for all entangled states. The question whether $E_D>0$ for all entangled states was answered negatively. Indeed, Horodecki et.\ al.\ showed that there exist entangled states from which no entanglement can be distilled at all \cite{HHH98}. When $E_D>0$ we call the state distillable, otherwise it is called a bound entangled state.

The question was then to classify all bound entangled states. This is also known as `the distillability problem' and is the main focus of this paper (see problem 2 in \cite{KW05}). We start off by reviewing what is known. The next theorem is crucial.

\begin{theorem}[Horodecki et al.\ \cite{HHH97, HHH98}]
\label{funthe}
(i) All entangled two qubit states are distillable.

(ii) An arbitrary bipartite state $\rho$ acting on ${\cal H}_A \otimes {\cal H}_B$ is distillable if and only if there exist projectors $P:{\cal H}^{\otimes n}_A \rightarrow {\cal H}_2 $ and  $Q:{\cal H}^{\otimes n}_B \rightarrow {\cal H}_2 $ and a number $n$, such that the state 
\begin{align}
\label{pqpq}
\rho'=(P\otimes Q)\rho^{\otimes n} (P\otimes Q)^{\dagger}
\end{align}
is entangled. Since $\rho'$ is a state acting on ${\cal H}_2 \otimes {\cal H}_2$, this means that $\rho'$ needs to have a negative partial transposition: $(\rho')^{T_B}<0$. 
\end{theorem}

If such an $n$ exists we call $\rho$ pseudo-$n$-copy distillable or in short $n$-distillable. The prefix $pseudo$ reflects the fact that if we project upon such a subspace we are only half way through our distillation process. Indeed, in the next step we would like to repeat this procedure $m$ times on batches of $n$ copies of $\rho$, giving us $m$ copies of the qubit pair $\rho'$. Finally we can use existing protocols to extract maximally entangled singlets from $\rho'^{\otimes m}$ \cite{BBPSSW96,BDSW96}.

The theorem is equivalent \cite{DCLB99} to saying that if a state $\rho$ is distillable then we can find a Schmidt rank two vector $\psi$ and a number $n$ such that $\la \psi | (\rho^{\otimes n})^{T_B} |\psi\ra<0$. From this it follows that states with a positive partial transposition (PPT) can never be distilled (see \cite{Alb01} for a direct proof). Usually the term `bound entangled states' is therefore associated with entangled PPT states. 
The question remains whether there exist bound entangled states with a negative partial transposition (NPT).
 This problem can be reduced to the question whether all entangled Werner states are distillable \cite{HH97, Werner89, Alb01}. Recall that a Werner state acting on ${\cal H}_A\otimes {\cal H}_B \cong \mathbb{C}^d \otimes \mathbb{C}^d$ can be written as 
$$
\rho_W = \frac{1}{d^2+\beta d}(\ido +\beta F), \qquad -1\leq \beta \leq 1.
$$
Here $F=\sum_{ij} |ij\ra\la ji|$ denotes the swap or flip operator. These states are entangled when $\Tr (\rho_W F)<0$ or $\beta < -1/d$. A prominent property of these states is that they are the only states which are invariant under local unitary transformations of the form $U\otimes U$. Hence, any state $\rho$ can be transformed into a Werner state by applying the so called twirl operation:
$$
\rho_W = \int dU U\otimes U \rho (U\otimes U)^\dagger,
$$
where the integral is with respect to the Haar measure on $U(d)$.
Note that this transformation leaves the expectation value $\Tr (\rho_W F)=\Tr (\rho F)$ invariant.
Now it is easy to prove that one can transform any NPT state $\rho$ to an NPT state $\rho'$ such that $\Tr (\rho' F)<0$. Applying the twirl then gives an entangled Werner state.

The distillability of the Werner states has been studied in two papers \cite{DCLB99, DSSTT00}. The authors were able to show that they are distillable when $\beta<-1/2$ and $n$-copy undistillable for $\beta>-1/d+\epsilon_n$. Unfortunately, the range $\epsilon_n$ goes to zero as $n$ goes to infinity. It is however conjectured that the Werner states are undistillable for all $\beta\geq -1/2$. 
An important result in this context was obtained by Watrous \cite{Watrous03} who constructed a one parameter set of distillable states which are $n$-copy undistillable in some range. Supporting evidence in favour of the conjecture was provided \cite{DCLB99} in the form of numerical evidence for $2$ and $3$ copies for $d=3$. 
Apart from intrinsic importance of the conjecture, affirmation would imply non-additivity and non-convexity of bipartite distillable entanglement \cite{SST01}. The problem also has non-trivial consequences on the theory of positive maps \cite{DSSTT00}.

A remarkable effect in the context of distillation, is activation of PPT bound entanglement. It has been proven that every state becomes 1-distillable by adding a PPT bound entangled state \cite{EVWW01, VW02} (see also \cite{CDKL01}). Conversely it has been shown that for every PPT state there exists a 1-undistillable state, such that taken together one obtains a 1-distillable state \cite{Masanes05}. 

In the next section we will discuss the distillability properties of certain class of highly symmetric states, which includes the Watrous states and two copies of the Werner states. We will come back to this set of states repeatedly in the rest of the paper. 
Section III is the main part of this paper, here we reformulate the distillability problem first as a special instance of the Schmidt number problem, and using similar techniques as an instance of the separability problem. We suggest and discuss several approaches to tackle this specific separability problem.
In section IV we outline our numerical method for detecting distillable states, we give a numerical estimate of the volume of distillable states for low dimensions and we provide strong evidence that all 1-undistillable Werner states are also 4-undistillable. In the Appendix we present a one parameter set of states which appears to exhibit all forms of distillability. In most of this paper we will omit the normalisation of density operators as they are not relevant to us.

\section{The $UUVVF$-invariant states}
To start we will briefly recapitulate some properties of the so-called local symmetry groups and 
states invariant under such groups. For an excellent overview with plenty of examples, the reader is referred to Ref.\ \cite{VW01}.

Let $G_0$ be a subgroup of the unitaries, or possibly the whole group of the unitaries.
Then we can define the group $G$ of the unitaries consisting of all pairs of the form $U \otimes U'$ acting on a Hilbert space ${\cal H}={\cal H}_A\otimes {\cal H}_B$, where $U\in G_0$ and $U'$ is some given function of $U$.
The set of bipartite states left invariant by $G$ is just the intersection of the state space with the commutant of the group $G$.  Generally speaking, choosing $G_0$ sufficiently large, there will exist a finite basis of operators spanning the commutant. The set of bipartite states left invariant by the group $G$ will also be denoted as $UU'$-invariant instead of $G$-invariant. An arbitrary state can projected onto an $UU'$-invariant state by twirling it
$$
{\cal T}_G(\rho) = \int dU U\otimes U' \rho (U\otimes U')^\dagger,
$$
here the integral is performed according to the Haar measure on $G_0$.
In the introduction we have seen an example of such a local symmetric set of states, namely the Werner states, which are $UU$-invariant and spanned by $\ido$ and $F$. The so called isotropic states \cite{HH97, Rains98} are $UU^*$-invariant and a basis is given by $\ido$ and $P=1/d\sum_{ij} |ii\ra\la jj|$. 

From these two basic symmetry groups one can generate others by considering tensor products. Consider the case where we have two symmetry groups $G=\{U\otimes U'\}$ and $K=\{V\otimes V'\}$ acting respectively on ${\cal H}_1={\cal H}_{A}\otimes {\cal H}_{B}$ and ${\cal H}_2={\cal H'}_{A}\otimes {\cal H'}_{B}$. Let $B_G$ and $B_K$ be a basis for the $UU'$- and the $VV'$-invariant states respectively. 
Then a basis for the $UU'VV'$-invariant states acting on ${\cal H}_1\otimes {\cal H}_2$ will be given by $B_G \otimes B_K$. In what follows we will number the systems belonging to party $A$ with odd numbers and party $B$ with even numbers. As an example, a basis for the $UUVV$-invariant states is given by the operators $\{\ido_{12}\otimes \ido_{34}, F_{12} \otimes \ido_{34}, \ido_{12}\otimes F_{34}, F_{12}\otimes F_{34} \}$.
Imposing the extra condition that the coefficients of $F_{12} \otimes \ido_{34}$ and $\ido_{12}\otimes F_{34}$ should be the same, we end up with the so-called $UUVVF$-invariant states as introduced in \cite{VW01}, where it was used as a counterexample to the additivity conjecture for the relative entropy of entanglement.

The set of $UUVVF$-invariant states contains the Werner states and the Watrous states and hence is also ideal for the study of distillability. A convenient parametrisation is given by
\begin{align*}
\rho=\ido_{12}\otimes\ido_{34}+\frac{\epsilon d-1}{d}(\ido_{12}\otimes F_{34} + 
 F_{12} \otimes \ido_{34}) + \frac{1-2\epsilon d+\delta d^2}{d^2}F_{12}\otimes F_{34}.
\end{align*}
The set of density operators is restricted by the following inequalities
\begin{align*}
(d-1)^2+2\epsilon d(d-1)+\delta d^2 & \geq 0, \\
d^2-1+2\epsilon d-\delta d^2 & \geq 0,\\
(d+1)^2-2\epsilon d (d+1)+\delta d^2 & \geq 0.
\end{align*}
This set of states includes the Werner \cite{Werner89} states (2 pairs in $d\otimes d$ for $\delta =\epsilon^2$ and 1 pair in $d^2\otimes d^2$ for $\epsilon=1/d$) and the Watrous \cite{Watrous03} states ($1-2\epsilon d+\delta d^2=d^2$). The PPT states are just the separable states (see \cite{VW01}). 
The states are entangled for $\epsilon<0$ or $\delta<0$ which follows from 
\begin{align}
\label{ptbrho}
\rho^{T_B}=Q_{12}\otimes Q_{34} + d\epsilon(P_{12}\otimes Q_{34} + Q_{12}\otimes P_{34}) +   \delta d^2 P_{12}\otimes P_{34},
\end{align}
with $Q=\ido-P$. 

\begin{figure}
\includegraphics[height=8cm]{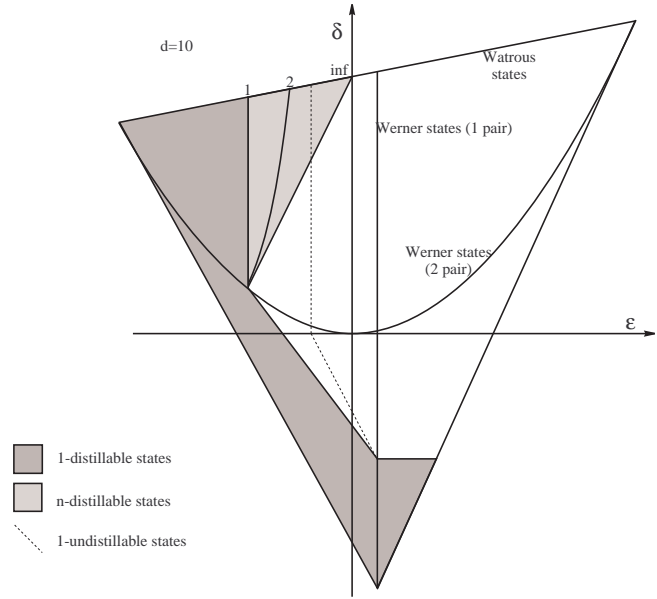}
\caption{$UUVVF$-invariant states. All states satisfying $\epsilon\geq 0$ and $\delta\geq 0$ are separable. Shaded areas mark out distillable states ($1$,$2$ and $\infty$- copy distillable).} 
\label{fig1}
\end{figure}

Let us now investigate the distillation properties of these states. To prove distillability, all we need to do is to find Schmidt rank two vectors $\psi$ such that 
$\la \psi | \rho^{T_B}|\psi\ra<0$. We present three such vectors, which we conjecture to detect all 1-distillable states:
\begin{align}
|\psi\ra_A&=|00\ra_{12} \otimes (|00\ra+|11\ra)_{34}, \nonumber\\
|\psi\ra_B&=|01\ra_{12} \otimes (|00\ra+|11\ra)_{34}, \label{ABC}\\
|\psi\ra_C&=\sum_{ij}|ii\ra_A|jj\ra_B+|i(i+1)\ra_A|j(j+1)\ra_B\nonumber.
\end{align}
which detect distillability for the states respectively satisfying
\begin{gather*}
d^2+3d(\epsilon d-1)+2(1-2\epsilon d +\delta d^2)<0,\\
\epsilon <\frac{1}{d}-\frac{1}{2}, \\
\delta <\frac{1}{d^2}-\frac{1}{2}.
\end{gather*}
This set of states is shaded dark grey in Fig.~\ref{fig1}.

Now let us derive results in the other direction, namely which states are undistillable? First let us show that we can find NPT states which are $n$-undistillable for arbitrary $n$. We will use the following inequalities derived in \cite{DCLB99}
\begin{itemize}
\item $\la \psi | \ido^{\otimes N-k}\otimes P^{\otimes k} | \psi \ra \leq  \frac{2}{d^k}$,
\item $\la \psi |  Q^{\otimes N-k} \otimes P^{\otimes k} | \psi \ra \leq  \frac{2}{d^k}$,
\item $\la \psi | Q^{\otimes N} | \psi \ra \geq  (1-\frac{2}{d})^N$,
\end{itemize}
for any $|\psi\ra$ with Schmidt rank two. Now from (\ref{ptbrho}) it follows that
 all potential negative terms contain at least one $P$ term. But we will also have a term $\la \Psi | Q^{\otimes 2N} | \Psi \ra \geq  (1-\frac{2}{d})^{2N}$, and this term can always dominate when we choose $\epsilon$ and $\delta$ small enough. Thus for each $n$, as long as we choose $\epsilon$ \emph{or} $\delta$ small enough (but one or both negative), we obtain an $n$-undistillable NPT state. 

Next, let us derive states that are definitely 1-undistillable. From (\ref{ptbrho}) and the given inequalities it follows straightforwardly that the states satisfying
$$
(1-2/d)^2+\min(4\epsilon,0)+\min(2\delta,0)\geq 0,
$$
are 1-undistillable. Note that this set does not touch the set of 1-distillable states. The reason for this is that although the inequalities are sharp, the sum of such inequalities are not. In some regions we can get a better bound by rewriting (\ref{ptbrho}) as
\begin{align*}
\rho^{T_B}=\ido_{12}\otimes \ido_{34} + (\epsilon d-1)(P_{12}\otimes Q_{34} + Q_{12}\otimes P_{34}) + (\delta d^2-1) P_{12}\otimes P_{34}, 
\end{align*}
and we find that all states such that
$$
d^2+\min(4d(\epsilon d-1),0)+\min(2(\delta d^2 -1),0) \geq 0
$$
are 1-undistillable. This second set of states is depicted in Fig.~\ref{fig1}.

To conclude our discussion of the distillation properties we will discuss the Watrous states \cite{Watrous03} of which the general form is given by
\begin{align*}
\rho=\ido_{1,2}\otimes\ido_{3,4}+\frac{\epsilon d-1}{d}(\ido_{1,2}\otimes F_{3,4} + 
 F_{1,2} \otimes \ido_{3,4}) + F_{1,2}\otimes F_{3,4},
\end{align*}
with $1+1/d>\epsilon >1/d-1$. The states are entangled if and only if $\epsilon<0$ and definitely 1-distillable if $\epsilon <1/d-1/2$. We will show that all entangled Watrous states are distillable. Suppose now we have two pairs of entangled Watrous states, the second pair having indices $5,6$ and $7,8$. Using the identities
\begin{align*}
\Tr((P_{1,5}\otimes P_{2,6})(\ido_{1,2}\otimes F_{5,6}))&=1/d, \\
\Tr((P_{1,5}\otimes P_{2,6})(F_{1,2}\otimes \ido_{5,6}))&=1/d, \\ 
\Tr((P_{1,5}\otimes P_{2,6})(F_{1,2}\otimes F_{5,6}))&=1,
\end{align*}
it can be verified that projecting upon $P_{1,5}\otimes P_{2,6}$ will yield a new Watrous state
$\rho'$ with parameter
$$
\epsilon'=\epsilon\Bigl(\frac{2(\epsilon d+d^2-1)}{d^2 \epsilon^2+d^2-1}\Bigr).
$$
Since $\epsilon'<\epsilon$ for $0>\epsilon>1/d-1$, the state will be more entangled than the state we started from for each $\epsilon<0$. One can repeat this protocol on many pairs until finally $\epsilon<1/d-1/2$, at which point we obtain a 1-distillable state. More generally the protocol can be applied to the whole set of states; a straightforward but tedious calculation leads to
\begin{align*}
\epsilon'&=\frac{\epsilon(d^2\delta + d^2-1)}{d^2 \epsilon^2 +d^2-1}, \\
\delta'&=\frac{\epsilon^2(d^2-1)+d^2\delta^2}{d^2 \epsilon^2 +d^2-1}.
\end{align*}
The states which are 2-distillable with this protocol are depicted in Fig.~\ref{fig1}.
Repeating the protocol recursively, it is not hard to show that all states satisfying
$$
\delta>\frac{3d^2+4d-8}{2d(d-2)}\epsilon+1-\frac{1}{d^2},
$$
are distillable (see again Fig.~\ref{fig1}). In section IV.A we will give evidence that all the other states are probably NPT undistillable, but this of course this awaits an analytical proof.

\section{A positive approach to the distillability problem}
As mentioned in the introduction, the distillability conjecture is equivalent to the statement that 
there exists no Schmidt rank two vector $|\psi\ra$ such that
$$
\la \psi | {\rho_W^{T_B {\otimes n}}} |\psi\ra < 0
$$
for all $n\geq 1$ and $\rho_W^{T_B} = \ido -\frac{d}{2} P$. In other words, it seems like affirmation of the conjecture would have to be in the form of an impossibility proof as opposed to a constructive proof.
 We will reformulate this conjecture in a more tractable form, namely as a special instance of the separability problem, for which a large number of tools are present.  As a steppingstone we first show how to translate the distillability problem into the problem of detecting Schmidt number 3. In the next subsection we then reformulate it as a separability problem.

\subsection{As a Schmidt number problem}
The Schmidt number of a quantum state has been introduced in \cite{TH00} as a generalisation of the Schmidt rank of a pure state. The Schmidt number of a state $\rho$ is defined as the smallest number $n$ such that $\rho$ can be written as a convex combination of pure states with Schmidt rank $n$. 
Thus, separable states have Schmidt number one and entangled states have a Schmidt number larger than one. The problem of detecting the Schmidt number of a state has received little attention until now. A notable exception is Ref.\ \cite{SBL00} which developed the notion of Schmidt-number witnesses. When a state is 1-undistillable we have that  $\Tr (|\psi \ra\la \psi | \rho^{T_B})>0$ for all Schmidt rank two states $\psi$. Thus for every NPT 1-undistillable state $\rho$, $\rho^{T_B}$ is positive on Schmidt rank two states and is thus a Schmidt number 3 witness.
Analogously, 1-distillable states will give rise to Schmidt number 2 witnesses \cite{SBL00}. 
The symmetry of the Werner states allows for a dual approach, reformulating the problem of proving that a certain operator is a Schmidt number 2 (three) witness into the problem of detecting the Schmidt number 3 of a certain class of states. We will do so for one, two and $n$ copies of the Werner states. 

Let us start with one copy of $\rho_W$, for which the answer is known. As $\rho_W$ belongs to the set of $UU$-invariant states, $\rho_W^{T_B}$ will belong to the set of $UU^*$-invariant states and hence will be invariant under the $UU^*$-twirl: ${\cal T}_{UU^*}(\rho_W^{T_B})=\rho_W^{T_B}$. From this follows that 
$$
\la \psi | \rho_W^{T_B} |\psi\ra = \Tr(P_\psi \rho_W^{T_B})=\Tr( {\cal T}_{UU^*}(P_\psi) \rho_W^{T_B}),
$$
here ${\cal T}_{UU^*}(P_\psi)$ is the operator $P_{\psi}=|\psi\ra\la\psi|$ after application of the $UU^*$-twirl. Thus we do not need to check over the whole set of Schmidt rank two vectors, but instead over the restricted set of $UU^*$-invariant states with Schmidt number 2. The Schmidt number of the $UU^*$-invariant states or the isotropic states is well known \cite{TH00, SBL00}. 
If we parametrise the isotropic states as $\rho_\alpha=\ido+\alpha P$ then $\rho_\alpha$ has Schmidt number $k$ when
$$
\alpha \leq \frac{d(kd-1)}{d-k}.
$$
From this follows that $\rho_W= \ido +\beta F$ is one-distillable if and only if $\Tr(\rho_W^{T_B} \rho_\alpha)<0$, with $\alpha=\frac{d(2d-1)}{d-2}$. Going through the algebra we recover that all Werner states $\rho_\beta$ with $\beta<-1/2$ are 1-distillable.

\begin{figure}
\includegraphics[height=8cm]{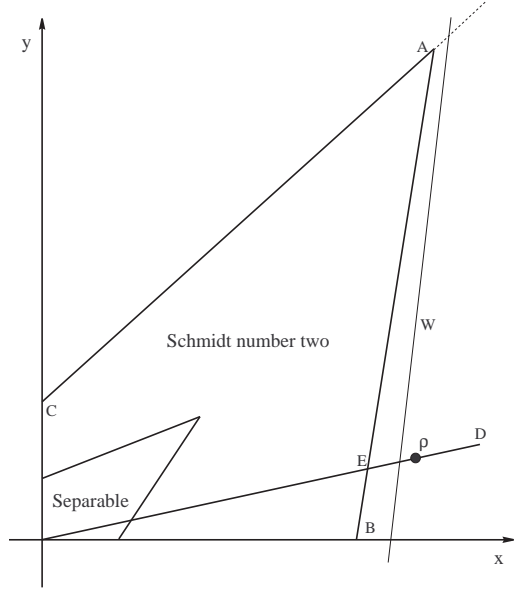}
\caption{$UU^*VV^*F$-invariant states.} 
\label{uuvvs}
\end{figure}

Let us now look at two copies of the Werner states $\rho_W^{\otimes 2}$; as pointed out before, these states belong to the larger class of the $UUVVF$-invariant states. Thus the relevant dual set is the set of the $UU^*VV^*F$-invariant states. For convenience we write the operators in the order of the indices 1,2,3,4 and omit these indices. With this in mind we can parametrise the $UU^*VV^*F$-invariant states as
$$
\rho=Q\otimes Q +x (Q\otimes P + P\otimes Q) +y P \otimes P,
$$
with $x,y>0$. These states are separable for $d^2-2d(x-1)+(1-2x+y)\leq 0$ and $d^2-(1-2x+y)\leq 0$ as depicted in Fig.~\ref{uuvvs}. The set of Schmidt number 2 states contains at least the convex hull of the points 
\begin{align*}
A&=\biggl(\frac{(3d-4)(d+1)}{2d-4},\frac{2(d+1)^2(d-1)}{d-2} \biggr) \\
B&=\biggl(\frac{d^2-1}{d-2},0\biggr)\\
C&=\biggl(0,\frac{2(d^2-1)^2}{d^2-2}\biggr)
\end{align*}
obtained by twirling the Schmidt rank two vectors from equation~(\ref{ABC}). All the states lying above the $CA$-line have a Schmidt number larger than two. This follows from the fact (see section II) that the operator $\ido \otimes \ido - \frac{d^2}{2} P\otimes P$ is positive on Schmidt number 2 states. If the Werner states are 2-undistillable then also all the states lying to the right of the $AB$-line have Schmidt number at least three. This follows easily by evaluating the expectation value of $\rho$ at $(\ido  -\frac{d}{2}P)^{\otimes 2}$. 
It is important to note that it is sufficient to show that all the states $\rho$ on for instance the $ED$-segment (see Fig.~\ref{uuvvs}) have Schmidt number 3.
This can be seen as follows: let $\rho$ approach $E$, and suppose we can show that each $\rho$ has Schmidt number 3. This would imply the existence of a hyperplane $W$ separating $\rho$ from the set of Schmidt number 2 states. For $\rho$ arbitrarily close to $E$ this hyperplane would be parallel to the $AB$-segment, otherwise cutting it. Therefore this would show that all the states lying to the right of the $AB$-hyperplane have Schmidt number at least three.

For the general case we need to consider the set of $U_1U^*_1\cdots U_nU^*_nF$-invariant states. Here the subindex in $U_i$ refers to the subsystem the unitary operator is acting and the $F$ denotes any permutation of the subsystems. In what follows we will call these states
$U_iU^*_iF$-invariant states. We parametrise them as
\begin{align*}
\rho={\tilde Q}^{\otimes n}+x_1(P\otimes {\tilde Q}^{\otimes n-1} + {\tilde Q}\otimes P\otimes {\tilde Q}^{\otimes n-2}+\cdots)+ \\
 x_2(P^{\otimes 2}\otimes {\tilde Q}^{\otimes n-2} +  P\otimes {\tilde Q}\otimes P \otimes {\tilde Q}^{\otimes n-2}+\cdots) + \cdots + x_n P^{\otimes n},
\end{align*}
where we have found it convenient to now use the normalised ${\tilde Q}=Q/(d^2-1)$. The relevant hyperplane in this case is given by 
$$
\Tr\Bigl[\Bigl(Q+\Bigl(1-\frac{d}{2}\Bigr)P\Bigr)^{\otimes n} \rho\Bigr]=1+\sum_{i=1}^n \binom{n}{i}  \Bigl ( 1-\frac{d}{2}\Bigr)^i x_i=0.
$$
We can also write this as $1+\sum_{i=1}^n \binom{n}{i} (-1)^i {\tilde x}_i=0$ with ${\tilde x}_i= \bigl(\frac{d}{2}-1\bigr)^i x_i$. From now on we will continue to work in these normalised variables. Next we will generalise the idea developed for two copies. First we need to show that the hyperplane is spanned by Schmidt number 2 states. Then in order to check distillability, it will be enough to find the boundary between two and three Schmidt number along a line from the origin to an interior point of the points spanning the hyperplane.

An independent set of Schmidt number 2 states spanning the hyperplane is easily obtained as follows (compare to the case for two copies).
Let $|\psi\ra_1=\frac{1}{\sqrt{2}} |01\ra^{\otimes n-1} \otimes (|00\ra+|11\ra)$. Twirling this state will yield an $U_iU^*_iF$-invariant state with coordinates $(\frac{1}{n},0,\cdots,0)$. Similarly, twirling $|\psi\ra_2=\frac{1}{\sqrt{2}} |00\ra\otimes |01\ra^{\otimes n-2} \otimes (|00\ra+|11\ra)$ will yield an  $U_iU^*_iF$-invariant state with coordinates $({\tilde x}_1\neq 0,{\tilde x}_2\neq 0,0,\cdots,0)$. 
In general 
$$
|\psi\ra_k=\frac{1}{\sqrt{2}} |00\ra^{\otimes k} \otimes |01\ra^{\otimes n-k-1} \otimes (|00\ra+|11\ra)
$$
will yield a state with coordinates ${\tilde x}_i \neq 0$ for $i\leq k+1$ ${\tilde x}_i = 0$ for $i> k+1$. It is evident that all points will lie on the hyperplane and that they form an independent set, spanning the hyperplane. An interior point can for instance be obtained from the first point, as $(\frac{1}{n}+\epsilon(1-\frac{1}{n}),\epsilon,\cdots,\epsilon)$ for sufficiently small $\epsilon$. It can readily be verified that this point belongs to the hyperplane by using the identity $\sum_{i=2}^n \binom{n}{i}(-1)^i=n-1$.

Thus $n$-undistillability of the Werner states beyond the 1-distillability boundary is equivalent to the statement that the $U_iU^*_iF$-invariant state with coordinates $(\frac{1}{n}+\epsilon(1-\frac{1}{n})+\delta,\epsilon,\cdots,\epsilon)$ has Schmidt number 3 for $\epsilon>0$ small enough and all $\delta >0$.

\subsection{As a separability problem}
This section contains the main result of this paper, namely the casting of the distillability problem as a special instance of the separability problem. An important tool in this, is the following result.
\begin{theorem}[Kraus, Lewenstein and Cirac \cite{KLC01}]
\label{tttt}
Let $P_2$ be the projector onto a maximally entangled state acting on ${\cal H}_1={\cal H}_{A_1} \otimes {\cal H}_{B_1}={\mathbb C}^2 \otimes {\mathbb C}^2$. Then for an arbitrary operator $X$ acting on ${\cal H}_2={\cal H}_{A_2} \otimes {\cal H}_{B_2}$ we can define
\begin{align}
\label{waws}
W_X=P_2 \otimes X^{T_B}.
\end{align}
A state $\rho$ acting on  ${\cal H}_2$ is $n$-undistillable if and only if $W_{\rho^{\otimes n}}$ is an entanglement witness.
\end{theorem}
This theorem can readily be seen from the following lemma
\begin{lem}
\label{lemmm}
Let $\sigma$ be a positive operator with Schmidt number $N\geq 1$ acting on ${\cal H}_1={\cal H}_{A_1} \otimes {\cal H}_{B_1}$ and let $\eta$ be an operator acting on ${\cal H}_2={\cal H}_{A_2} \otimes {\cal H}_{B_2}$ positive on states with Schmidt number $KN$. Then the operator $\sigma \otimes \eta$ acting on ${\cal H}_{1} \otimes {\cal H}_{2}\cong {\cal H}_{A} \otimes {\cal H}_{B}$ is positive on states with Schmidt number $K$.
\end{lem}
\begin{proof}
It is clear that it is sufficient to prove the lemma for pure states $\sigma=|\phi\ra\la\phi|$. So let 
$$
|\phi\ra=\sum_{i=1}^N \phi_i |a_i\ra_{A_1}|b_i\ra_{B_1},
$$
and take an arbitrary Schmidt rank $K$ state 
$$
|\psi\ra=\sum_{j=1}^K \psi_j |e_j\ra_{A}|f_j\ra_{B}.
$$
Then we need to prove that $\Tr(|\psi\ra\la\psi| \sigma \otimes \eta)>0$. This trace operation can be composed of tracing out the first pair, and then the second, as $\Tr(\cdot)=\Tr_2(\Tr_1(\cdot))$. Then making use of the identity $\Tr_1(C(A_1\otimes B_2))=\Tr_1(C(A_1\otimes \ido_2))B_2$ we have
$$
\Tr(|\psi\ra\la\psi| \sigma \otimes \eta)= \Tr_2( \Tr_1(|\psi\ra\la\psi| (\sigma\otimes \ido_2)) \eta)
$$
Now $\Tr_1(|\psi\ra\la\psi| (\sigma\otimes \ido_2))$ is the projector onto the pure Schmidt rank $KN$ state $|\gamma\ra_2= \la\phi|\psi\ra = \sum_{ij} \phi_i\psi_j \la a_i |e_j \ra\la b_i |f_j \ra \in {\cal H}_2$ from which the lemma follows.
\end{proof}
Special cases of this lemma have appeared in the literature over the years \cite{KLC01, DCLB99,VW02, HBLS04}. Next we will apply Theorem~\ref{tttt} to the Werner states for one and $n$ copies and making use again of the local symmetry, we will present a dual positive formulation of the conjecture. The analysis will be very analogous to the reformulation as a Schmidt number problem and a continuous comparison of this section with the previous is very useful.

\begin{figure}
\includegraphics[height=8.5cm]{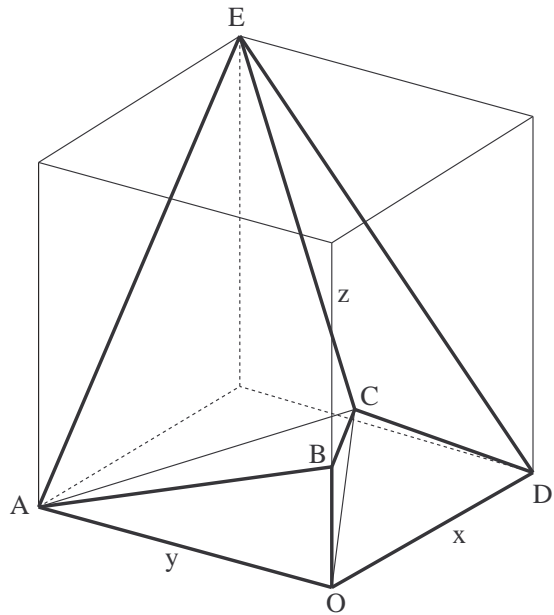}
\caption{The state space of the $UU^*VV^*$-invariant states. The point C lies in the $xz$-plane such that the points A, B, C and E lie in the same plane.} 
\label{threed}
\end{figure}

For one pair, we need to prove that $P_2\otimes (\ido-\frac{d}{2}P)$ is an entanglement witness. As before, it will be sufficient to characterize the subset of the separable states of the general $UU^*VV^*$ invariant states. Here $U$ acts on a two-dimensional Hilbert space. We parametrize the $UU^*VV^*$-invariant states as
$$
\rho=\tilde Q_2\otimes\tilde Q+xP_2\otimes \tilde Q+y\tilde Q_2\otimes P+zP_2\otimes P,
$$
with $x,y,z\geq 0$. The separable states are a subset of the states with positive partial transpose, which satisfy the inequalities
\begin{align*}
3z\leq\frac{1+3x}{d-1}-y & \qquad & \text{$ABCE$-plane}\\
z\leq\frac{1-x}{d+1}+y  & \qquad & \text{$ECD$-plane} \\
z\geq \frac{x-1}{d-1}+y & \qquad & \text{$EAD$-plane}.
\end{align*}
Thus the states with positive partial transition are contained in the polyhedron spanned by the points OABCDE as in Fig~\ref{threed}. By twirling the pure separable states $|0101\ra, |0100\ra, \sum_{i,j=0}^1|ijij\ra, |0001\ra$ and $|0000\ra$ one obtains the states represented respectively by the points $O,A,C,D$ and $E$. Note that $B$ is not in this list. Indeed, as 
we know that $P_2\otimes (\ido-\frac{d}{2}P)$ is an entanglement witness, all states satisfying $z>2x/(d-2)$ are entangled. In particular the states in the tetrahedron spanned by the points $ABCO$ are PPT entangled. Conversely, knowing that the polyhedron $ABCO$ is PPT entangled immediately proves that $P_2\otimes (\ido-\frac{d}{2}P)$ is an entanglement witness. 

For two copies, it is sufficient to study the set of $UU^*(V_1V_1^*V_2V_2^*F)$-invariant states parametrised as
\begin{align*}
\rho= \tilde{Q}_2 \otimes(\tilde{Q}\otimes \tilde{Q} + y_1 (P \otimes \tilde{Q}+  \tilde{Q} \otimes P) +y_2 P\otimes P)+ P_2\otimes (x_0 \tilde{Q} \otimes \tilde{Q} + x_1 (P \otimes \tilde{Q}+  \tilde{Q} \otimes P) +x_2 P\otimes P)
\end{align*}
We will not attempt to completely classify the separable subset, instead it is enough to look at what happens in the neighbourhood of the hyperplane 
\begin{align*}
\Tr\Bigl[P_2\otimes \Bigl(Q+\Bigl(1-\frac{d}{2}\Bigr)P\Bigr)^{\otimes 2} \rho\Bigr]=
x_0-(d-2)x_1+\Bigl(\frac{d-2}{2}\Bigr)^2 x_2=0.
\end{align*}
We now show that this hyperplane is spanned by separable states. Consider the following pure separable states and their coordinates $(y_1,y_2,x_0,x_1,x_2)$ after action of the $UU^*(V_1V_1^*V_2V_2^*F)$ twirl:
\begin{align*}
|01\ra|01\ra|01\ra: & \quad (0,0,0,0,0) \\
|01\ra|00\ra|01\ra: & \quad \Biggl(\frac{1}{2(d-1)},0,0,0,0 \Biggr) \\
|01\ra|00\ra|00\ra: & \quad \Biggl(\frac{d-1}{d^2-d-1},\frac{1}{d^2-d-1},0,0,0 \Biggr) \\
\sum_{i,j=0}^1 |ij\ra|ij\ra|00\ra:&\quad \Biggl(\frac{1}{2(d-1)},0,\frac{d-2}{3d},\frac{3d-4}{6d(d-1)},\frac{2}{3d(d-1)}\Biggr) \\
\sum_{i,j=0}^1 |ij\ra|ij\ra|01\ra: &\quad \Biggl(0,0,\frac{d-2}{3d},\frac{1}{3d},0\Biggr).
\end{align*}
These coordinates can be verified after a tedious but straightforward calculation.

A point in the interior of the convex hull of these points can be obtained by averaging these coordinates. In this case one obtains the state with coordinates $x_0=2(d-2)/(15d)$, $x_1=(5d-6)/(30d(d-1))$, $x_2=2/(15 d(d-1))$, $y_1=d(2d-3)/(5(d-1)(d^2-d-1))$ and $y_2=1/(5(d^2-d-1))$. From this follows that the Werner states are 2-undistillable if and only if the $UU^*(V_1V_1^*V_2V_2^*F)$-invariant states with coordinates $(y_1,y_2,x_0,x_1+\epsilon,x_2)$ are entangled for all $\epsilon>0$.

Let us now move to $n$ copies. The relevant set of states is the set of $UU^*(V_iV^*_iF)$-invariant states
\begin{align*}
\rho={\tilde Q}_2 \otimes {\tilde Q}^{\otimes n} + 
{\tilde Q}_2 \otimes [  y_1(P\otimes {\tilde Q}^{\otimes n-1} + {\tilde Q}\otimes P\otimes {\tilde Q}^{\otimes n-2}+\cdots)+  y_2(P^{\otimes 2}\otimes {\tilde Q}^{\otimes n-2} +  \cdots)+\cdots  + y_n P^{\otimes n-1}]+ \\ 
P_2 \otimes [  x_0 {\tilde Q}^{\otimes n} + x_1(P\otimes {\tilde Q}^{\otimes n-1} + {\tilde Q}\otimes P\otimes {\tilde Q}^{\otimes n-2}+\cdots)+  \cdots + x_n P^{\otimes n}], \\
\end{align*}
with $x_i,y_i\geq 0$. The relevant hyperplane is given by
\begin{align*}
\Tr\Bigl[P_2\otimes \Bigl(Q+\Bigl(1-\frac{d}{2}\Bigr)P\Bigr)^{\otimes n} \rho\Bigr]= 
\sum_{i=1}^n \binom{n}{i}  \Bigl ( 1-\frac{d}{2}\Bigr)^i x_i=0.
\end{align*}
Renormalising $x_i$, this can be rewritten as $\sum_{i=1}^n \binom{n}{i} (-1)^i {\tilde x}_i=0$ with ${\tilde x}_i= \bigl(\frac{d}{2}-1\bigr)^i x_i$. Next we will show that this hyperplane touches the set of separable states by constructing a set of $2n+1$ separable states spanning the hyperplane.
The first $n+1$ states are obtained by twirling 
$$
|\psi\ra_k=|01\ra |00\ra^{\otimes k}  |01\ra^{\otimes n-k},
$$
 for $k=0,\cdots, n$. The twirled state will be  $UU^*(V_iV^*_iF)$-invariant and will satisfy ${\tilde x}_i = 0$ for all $i$ and ${\tilde y}_j \neq 0$ for $j\leq k$ and  ${\tilde y}_j = 0$ for $j>k$. 
The last $n$ states are obtained by twirling 
$$
|\psi\ra_k=\sum_{i,j=0}^1|ij\ra|ij\ra |00\ra^{\otimes n-k-1}  |01\ra^{\otimes k},
$$
for $k=0,\cdots, n-1$. The twirled state will be  $UU^*(V_iV^*_iF)$-invariant and the $x$ coordinates will satisfy $x_j=0$ for $j> n-k$ and $x_j\neq 0$ for $j\leq n-k$. Therefore the coordinates of the $2n+1$ states are linearly independent and the choice of $|\psi\ra_k$ guarantees that the states will lie in the hyperplane.
An interior point in the convex hull of these $2n+1$ points can be obtained by choosing $\tilde x_0=\epsilon$ and $\tilde y_j=\tilde x_j=\epsilon$, for $\epsilon$ sufficiently small. One verifies that this point belongs to the hyperplane using the identity $\sum_{i=0}^{n-1}\binom{n}{i} (-1)^i =-1$.
From this follows that the conjecture is equivalent to the statement that the $UU^*(V_iV^*_iF)$-invariant states with coordinates $\tilde y_i=\epsilon$ , $\tilde x_i=\epsilon$ and $\tilde x_1=\epsilon+\delta$, where $j=1,\cdots,n$ and $i=0,2,\cdots,n$, are entangled for $\epsilon>0$ small enough and all $\delta >0$.

\subsection{Discussion}
The general separability problem has been proven to be NP-hard \cite{Gurvits03}. However, above we showed that the distillability problem can be reformulated as the question of entanglement of a \emph{particular} set of one parameter states. 
The first and by far still the most elegant tool for detecting entanglement is the partial  transposition criterion \cite{Peres93}. However, it is not useful to solve the dual entanglement problem, as it was proven in \cite{KLC01} that $W_{\rho^{\otimes n}_W}$ is a decomposable entanglement witness if and only if $\rho$ has positive partial transition. Hence $W_{\rho^{\otimes n}_W}$ will be either a non-decomposable witness or no entanglement witness at all depending on whether $\rho^{\otimes n}_W$ is 1-undistillable or 1-distillable. Thus in order to solve the dual problem, we will either need some powerful tool for detecting PPT-entanglement or a tool for proving separability. However, in the latter case, the original formulation stemming from Theorem~\ref{funthe} seems easier and for this purpose we present an efficient algorithm for detecting distillability in section~\ref{peasantsec}.

Probably the most powerful method for detecting entanglement is the complete family of separability criteria introduced by Doherty et.\ al.\ \cite{DPS01, DPS03}. Basically, their method relies on a hierarchical characterisation of separable states which they use to devise a computational algorithm for detecting entanglement. It has a number of very appealing features: 
(i) The set of criteria is complete, all entangled states are detected at some stage. 
(ii) The criteria can be cast into a semidefinite program which is a convex optimisation problem for which efficient algorithms exist. 
(iii) When a state is found entangled, their algorithm automatically yields an entanglement witness for that state. These entanglement witnesses turn out to be of a special form, namely such that after some manipulation the associated bihermitian form can be written as sums of squares. 
Bihermitian forms which can be written as sums of squares are canonical entanglement witnesses.
Therefore it is always possible to extract some analytical provable entanglement witness from the output of the algorithm. In principle therefore, the distillability problem can be solved for any number of copies using our dual formulation together with the algorithm associated with the complete family of separability criteria. 

Another numerical method based on semidefinite programming is the one introduced by Eisert et.\ al.\
\cite{EHGC04}. There the separability problem is cast as a global optimisation problem with polynomial constraints. Using the theory of semi-definite relaxations, a hierarchy of efficiently solvable approximations to the optimal solution is provided. Every entangled state is necessarily detected in some step of the hierarchy and since a global optimum will be found at some point, separability can also be detected. 
However, it should be noted that the reformulation of the distillability problem as a separability problem is unnecessary here, as their method allows for a simple way of checking whether or not an operator is an entanglement witness.
Using analogous techniques, it can be extended to testing whether or not an operator is a Schmidt number 3 witness \cite{Eisert}. 

The family of criteria introduced by Doherty et.\ al., however, has the advantage that an \emph{analytical} canonical entanglement witness can be extracted.
Unfortunately, we were unable to test either algorithm, as both methods seem only practical for low dimensional systems. The interesting case, two copies of the Werner states for $d=3$ could hence not be tested. It would be worth investigating whether these numerical methods could be simplified for the highly symmetrical states we are interested in.

Numerical solutions are one option, another possible approach would be one of a more indirect nature.
A powerful method for proving that a certain state is entangled is to show that, when shared by two parties, the state can enhance typical quantum operations such as teleportation or distillation. 
In particular, it was shown that PPT-preserving operations can be simulated using LOCC operations when both parties share PPT entangled states \cite{CDKL01}. The class of PPT-preserving operations is strictly larger than the LOCC class, so that we can expect it to do more. Recently it has been shown \cite{Masanes05} that \emph{every} entangled state can enhance the so called conclusive teleportation fidelity of some other state. It was also proven that for every PPT state $\sigma$ there is a 1-undistillable state $\rho$ such that $\rho\otimes \sigma$ is 1-distillable. 
These characterisations of entangled states seem very promising as a way of proving that a certain state is PPT entangled. 
In Ref.\ \cite{Ishizaka04} a class of PPT states was constructed which was shown to provide overall convertibility of pure entangled states. In particular it was shown to be able to increase the Schmidt number of a pure state. Now the PPT entangled states we obtained for one copy of the Werner states (tetrahedron ABCO in Fig.\ \ref{threed}) are of a similar form, and although not necessarily in an optimal way, they too can help increase the Schmidt number of a pure state.
Similar activation effects can be expected from the conjectured PPT states derived from two copies of the Werner states.

\section{The peasant's method}
\begin{figure}
\includegraphics[height=7.5cm]{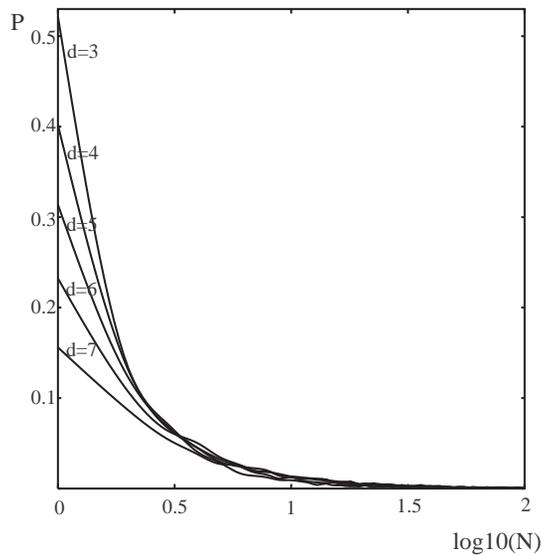}
\caption{The probability of detecting 1-distillability as a function of the test number for random states drawn from ${\cal D} \times {\cal U}$. For visual purposes only the first 100 tests are depicted and the points have been smoothened out to curves.} 
\label{probtest}
\end{figure}
\label{peasantsec}
In this section we will outline a powerful algorithm for the detection of distillable states. As $n$-distillability of $\rho$ is equivalent to 1-distillability of $\rho^{\otimes n}$, it is clear that we can confine ourselves to the study of 1-distillability. The basic problem we need to solve is the following minimisation over Schmidt rank two vectors:
\begin{align}
\label{bprob}
\min_{\psi \in SR2} \la \psi | \rho^{T_B}|\psi \ra  \overset{?}{<} 0.
\end{align}
The numerical method employed by D\"ur et al \cite{DCLB99} converts this problem to a minimisation of the minimum eigenvalue of certain matrices. Their method involves also calculating an inverse of a matrix in every step. The peasant's method \footnote{Name courtesy of Bram De Knock.} does not require this and is also conceptually simpler.

We can rewrite equation (\ref{bprob}) as 
$$
\min_{\psi \in SR2} \Tr (|\psi \ra\la \psi | \rho^{T_B})=\min_{D \in SR2} \Tr (D \rho^{T_B}).
$$
It is clear that the operators $D=|\psi \ra\la \psi |$ play the role of a distillability witness, in much the same way entanglement witnesses detect entanglement \cite{Terhal01}. 
So unlike in the entanglement problem, we have a complete characterisation of all distillability witnesses. Now it is well known that the positive map associated with an entanglement witness detects much more entanglement than the witness itself \cite{HHH96}. The positive maps $D(\cdot)$ associated with  $D=|\psi \ra\la \psi |$ can be chosen to act as 
$$
\label{todoeq}
D(\rho)=(\ido \otimes P) \rho^{T_B} (\ido \otimes P)^\dagger
$$
with $P=|0\ra\la a| + |1\ra\la b|$, with $\la b| a \ra=0$. This in effect is a generalisation of Theorem~\ref{funthe} stemming from the fact that in $n \otimes 2$ all NPT states are distillable (see also \cite{DCLB99, DSSTT00, Clarisse04}). Thus the problem is reduced to checking whether there exist vectors $|a\ra, |b\ra$ such that $\sigma$ has a negative eigenvalue. 
One way of doing this is to parametrize a countable subset of vectors which is dense within all vectors, such as the one introduced in Ref.\ \cite{HB04}. Explicitly their set takes the form $G=\{\sum \lambda_i |i\ra | (\lambda_1, \lambda_2,\cdots, \lambda_d ) \in \tilde G_{\mathbb{N}} \}$ with 
$$ 
\tilde G_N=\left\{\left(\frac{p_1}{q_1}e^{2\pi i \frac{r_1}{s_1}}, ,\frac{p_2}{q_2}e^{2\pi i \frac{r_2}{s_2}},\cdots, \sqrt{1-\sum_l \frac{p^2_l}{q^2_l}}e^{2\pi i \frac{r_d}{s_d}}\right)\right \}
$$
 for $ 0 < p_i\leq q_i \leq N; 0< r_i\leq s_i \leq N $.
Thus for every $N$, we can construct sets of pairs  $|a\ra$ and $|b\ra$, and taking $N$ increasingly large we will detect all 1-distillable states, except those arbitrary close to the boundary of the convex set of 1-undistillable states.

In practice two improvements can be made which greatly enhance the performance. First note that the countable subset above will yield vectors $|a\ra$ and $|b\ra$ not necessarily orthogonal. Furthermore, it is clear that such a countable subset will in general not pick vectors uniformly distributed according to the Haar measure. One way of overcoming this is to take for $|a\ra$ and $|b\ra$ two columns of a random unitary. Here orthogonality and uniformity are automatically guaranteed. The associated algorithm works very well for low dimensional density matrices. For higher dimensions, a local optimisation of the $|a\ra$ and $|b\ra$ yielding the minimum after a certain cut-off number of tests turns out to work well. 

We have checked the distillability of the $UUVVF$-invariant states over the complete range of parameters for 1 and 2 copies for $d=3$. We easily recovered the proposed boundaries for 1-distillability. For two copies, the states act on a $\mathbb{C}_{d^4} \otimes \mathbb{C}_{d^4}$ Hilbert space and numerical matrix manipulations in a space of this magnitude seem very hard. 
Fortunately, the states are very sparse and the peasant's method only requires minimisation of the minimum eigenvalue of a $2d^2\times 2d^2$ matrix. We were able to detect distillability for the Watrous states in $\epsilon>1/d-1/2$ readily and exhaustive testing suggest that also the proposed boundary for 2-distillability is correct. This provides very strong evidence that the Werner states are 4-undistillable for $\beta>-1/2$.

\subsection{Volume of 1-distillable states}
As an application of the peasant's method we will give a numerical estimate of the volume of 1-distillable states for low dimensional quantum states. A similar numerical estimate has been carried out for entangled states \cite{ZHSL98, Zyczkowski99}. When talking about volumes on the set of density operators, it is clear that the results will depend on the measure. We choose the measure applied in Ref.\ \cite{ZHSL98} as it seems very natural.

A general bipartite quantum state $\rho$ acting on ${\cal H}_A \otimes {\cal H}_B$ can be expanded by virtue of the spectral decomposition as
$$
\rho=UDU^\dagger,
$$
with $D=(d_{ii})$ diagonal and $U$ unitary. The measure we are going to use is the product measure ${\cal D} \times {\cal U}$. Here ${\cal D}$ represents the uniform distribution of the points on the manifold given by $\sum_i d_{ii}=1$. A simple method for generating such a distribution from  independent uniformly distributed random numbers chosen in the interval $(0,1)$ is outlined in Appendix A of Ref.\ \cite{ZHSL98}. Similarly, ${\cal U}$ is chosen to be the uniform measure on unitary matrices (the Haar measure). To generate random unitaries according to this measure one can use the algorithm from Ref.\ \cite{ZK94, PZK98} which relies on a decomposition of a general unitary in two-dimensional unitary transformations. A simpler method for generating Haar unitary matrices is as follows \cite{Stewart80, ER05}. Take a random matrix $A$, whose entries are complex numbers which are independently normally distributed with mean zero. The QR factorisation $A=QR$, such that $R$ has positive elements on the diagonal, then yields $Q$ distributed according to ${\cal U}$.

\begin{figure}
\includegraphics[height=7.5cm]{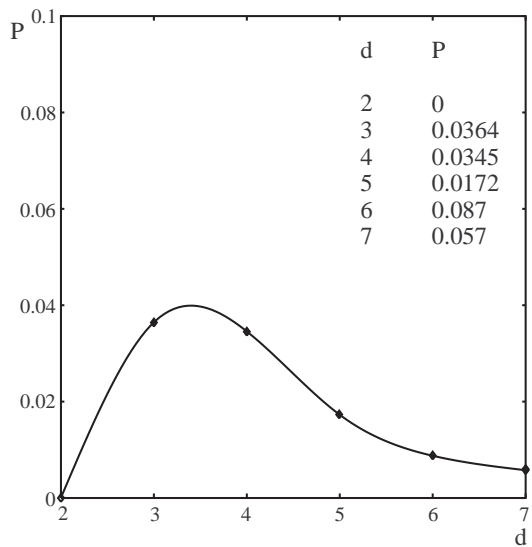}
\caption{Probability of a random NPT state drawn from ${\cal D} \times {\cal U}$ to be 1-undistillable. The curve has been drawn to guide the eye.} 
\label{volnpt}
\end{figure}

\begin{figure}
\includegraphics[height=7.5cm]{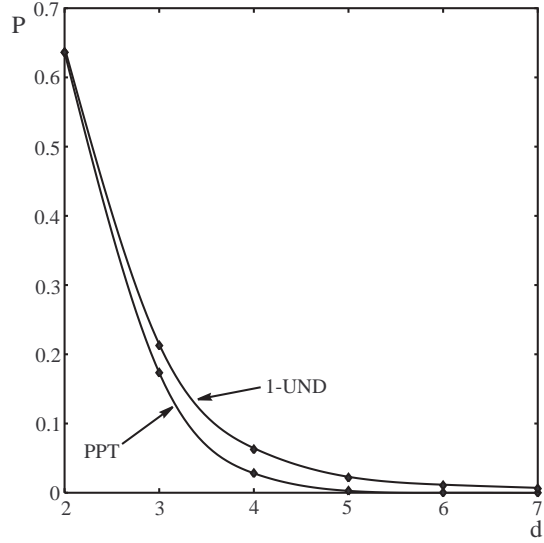}
\caption{Probability of a random state drawn from ${\cal D} \times {\cal U}$ to be 1-undistillable. } 
\label{volall}
\end{figure}

We have tested the peasant's method on $10^5$ density matrices acting on $\mathbb{C}^d \otimes \mathbb{C}^d$ for $d=3,4,5,6,7$. A striking feature is that the vast majority of the distillable states were detected in the first few tests. For example for $d=3$ about half of the NPT states are found to be distillable in the first test. The estimated probability of success as a function of the test number is displayed in Fig.~\ref{probtest}. For $d=7$ the probability of finding a distillable state in the first test is about $1/6$. This seems to suggest that the volume of 1-distillable states drops to zero for high dimensions. The opposite turns out to be more likely. 
In Fig.~\ref{volnpt} the probability of an NPT state being 1-undistillable is plotted versus the dimension $d$. To obtain sufficient precision we carried out $10^5$ random tests per state, and in addition $10^4d$ optimisation steps seeking for a local minimum. Of course, this method does not guarantee to detect every 1-distillable state, but we obtain an upper bound of the number of undistillable states. Note the distinct peak at $d=3,4$, the reason for this will be explained elsewhere \footnote{L.\ Clarisse, in preparation.}. In Fig.~\ref{volall} the same graph is drawn, but now PPT states are included. 

It is tempting to conclude from this numerical evidence that bound entangled states are primarily a phenomenon present in low dimensional quantum systems.
In high dimensional systems most undistillable states are therefore situated in the immediate neighbourhood of the set of separable states. Our results are consistent with the fact that bound entanglement for continuous variables is a rare phenomenon \cite{HCL01}. In particular it was shown that the subset of undistillable states is nowhere dense in the set of bipartite continuous variable states. From this it follows that the set of undistillable continuous variables states does not contain any open ball, an argument which was made explicit for separable states in \cite{ESP01}. Given an infinite dimensional separable state one can construct sequences of closer and closer states all of which are entangled. Following the same methods one can explicitly construct a distillable state in any $\epsilon$-neighbourhood in the trace norm of any state \cite{Eisert}. However, note that in Ref.\ \cite{HLW05} a parametrized family of measures on states was introduced, which in some region yields states primarily 1-undistillable.

\section{Conclusion}
The main result of this paper was that the distillability problem can be formulated as just a special instance of the separability problem. We have discussed several ways in tackling this separability problem which we believe merit further study. 
We have outlined an efficient numerical method for detecting distillability, and provided strong evidence that the distillability conjecture is valid at least up to 4 copies of the Werner states. The method was also used to make an estimate of the volume of 1-distillable states for $d=3,\ldots,7$.

\appendix
\section{The rainbow states}
Let us consider (we omit the indices $\{1,2\}$ and $\{3,4\}$)
\begin{align*}
\rho=\ido_m\otimes\ido_d+\frac{d\epsilon -1}{d}\ido_m\otimes F_d + \frac{m\epsilon -1}{m}F_m \otimes \ido_d +  \frac{1-(m+d)\epsilon+ d m \delta }{dm}F_m\otimes F_d.
\end{align*}
In what follows we will assume that $3\leq m<d$.
The set of density operators is restricted by the following inequalities
\begin{align*}
1+\delta+2\epsilon+\frac{1}{md}-(\epsilon+1)\frac{m+d}{md}& \geq  0\\
1-\frac{1}{md}+\epsilon\frac{m+d}{md}-\delta +\frac{1}{m}-\frac{1}{d}&\geq 0\\
1+\delta-2\epsilon+\frac{1}{md}+(1-\epsilon)\frac{m+d}{md}& \geq 0
\end{align*}
The partial transpose is given by
\begin{align*}
\rho^{T_B}=Q_m\otimes Q_d+m\epsilon P_m\otimes Q_d+ 
d\epsilon Q_m\otimes P_d+md\delta P_m\otimes P_d,
\end{align*}
and it is easy to see that the states are NPT iff $\epsilon<0$ or $\delta<0$. Now we will show that these states also include some PPT entangled ones. From Lemma~\ref{lemmm} and the results in \cite{TH00} it follows that
\begin{align*}
mP_m\otimes (\ido_d-\frac{d}{m}P_d)
\end{align*}
is an entanglement witness for $\rho^{T_B}$.  The partial transpose of an entanglement witness is again an entanglement witness:
\begin{align*}
F_m\otimes (\ido_d-\frac{1}{m}F_d).
\end{align*}
So that in addition apart from the states $\epsilon<0$ or $\delta<0$, the states that satisfy
\begin{align*}
\epsilon m^2 (d^2-1)+dm\delta(m-d)<0
\end{align*}
are entangled.

Let us now look at the distillation properties. Similar arguments to the ones we used for the $UUVVF$-invariant states apply here. For $\epsilon$ or $\delta$ sufficiently small, one can find $n$-undistillable states. The following vectors provide apparently the optimal boundaries for the 1-distillable states (we will only be interested in the states having $\delta>0$).

1.  $|\psi\ra=|00\ra_A|00\ra_B+|10\ra_A|10\ra_B$ gives rise to
\begin{align*}
2+2(d\epsilon-1)/d+4(m\epsilon-1)/m+ 4(1-(m+d)\epsilon+dm\delta)/(md)<0
\end{align*}

2.  $|\psi\ra=|00\ra_A|01\ra_B+|10\ra_A|11\ra_B$ gives rise to
\begin{align}
\epsilon <\frac{1}{m}-\frac{1}{2}.
\end{align}

\begin{figure}
\center{\includegraphics[height=8.5cm]{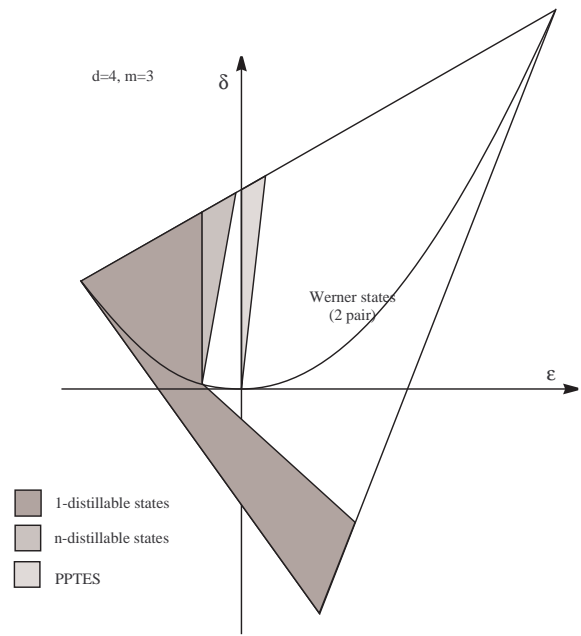}}
\caption{The top line of the triangle covers the whole spectrum of different types of distillability (the rainbow states)}
\label{fig2}
\end{figure}

Now let us take two pairs
\begin{align*}
\rho_1=\ido^{34}_m\otimes\ido^{12}_d+\frac{d\epsilon -1}{d}\ido^{34}_m\otimes F^{12}_d + \frac{m\epsilon -1}{m}F^{34}_m \otimes \ido^{12}_d + \frac{1-(m+d)\epsilon+ d m \delta }{dm}F^{34}_m\otimes F^{12}_d,
\end{align*}
and
\begin{align*}
\rho_2=\ido^{78}_m\otimes\ido^{56}_d+\frac{d\epsilon -1}{d}\ido^{78}_m\otimes F^{56}_d + \frac{m\epsilon -1}{m}F^{78}_m \otimes \ido^{56}_d  +\frac{1-(m+d)\epsilon+ d m \delta }{dm}F^{78}_m\otimes F^{56}_d .
\end{align*}
Taking both pairs together, and projecting upon $P^{1,5}_d\otimes P^{2,6}_d$, we end up with a $UUVVF$-invariant state in $m^2\otimes m^2$ with 
\begin{align*}
\epsilon'&=\frac{\epsilon(d^2\delta + d^2-1)}{d^2 \epsilon^2 +d^2-1} \\
\delta'&=\frac{\epsilon^2(d^2-1)+d^2\delta^2}{d^2 \epsilon^2 +d^2-1}.
\end{align*}
Now we know when these states can be distilled. The states distillable with this protocol are shown in Fig.~\ref{fig2}. One verifies that projecting upon $P^{1,5}_m\otimes P^{2,6}_m$ performs worse.
The set $1-\frac{1}{md}+\epsilon\frac{m+d}{md}-\delta +\frac{1}{m}-\frac{1}{d}=0$ contains states of all kinds: 1-distillable, $n$-undistillable but $n+1$-distillable, NPT undistillable (conjectured), PPT bound entangled and separable. 

\begin{acknowledgments}
This work was supported by a WW Smith Scholarship. I am very grateful to Anthony Sudbery for continuous support, advice and comments. I wish to thank Jens Eisert for sharing most useful comments and insights; and Andreas Winter for clarifying Ref.\ \cite{HLW05}. I would like to thank William Hall for comments on the final draft. I am grateful to Christine Aronsen Storebo for invaluable support.
\end{acknowledgments}

\bibliographystyle{apsrevl}
\bibliography{ent}
\end{document}